\documentclass[10pt]{amsart}
\usepackage{amsfonts}
\usepackage{amssymb}
\usepackage{amsmath,amsthm}
\usepackage{amscd}

\newcommand{\beq}{\begin{equation}}
\newcommand{\eeq}{\end{equation}}
\newcommand{\beqa}{\begin{eqnarray}}
\newcommand{\eeqa}{\end{eqnarray}}

\theoremstyle{definition}

\begin{document}


\begin{center}

{\Huge  Group entropies: from phase space geometry to entropy functionals via group theory}

\vspace{1cm}

Henrik Jeldtoft Jensen$^{1,2}$ and Piergiulio Tempesta$^3$

\vspace{.5cm}

$^{1}$ \quad Centre for Complexity Science and Department of Mathematics, Imperial College London, South Kensington Campus, SW7 2AZ, UK; h.jensen@imperial.ac.uk\\
$^{2}$ \quad Institute of Innovative Research, Tokyo Institute of Technology, 4259, Nagatsuta-cho, Yokohama 226-8502, Japan\\
$^{3}$ \quad Departamento de F\'{i}sica Te\'{o}rica II Universidad Complutense de Madrid
28040 - Madrid, Espan\~{a} and    Instituto de Ciencias Matematicas (ICMAT)
28049 - Madrid, Espan\~{a}; p.tempesta@fis.ucm.es
\vspace{1cm}

{\Large Abstract}\\
The entropy of Boltzmann-Gibbs, as proved by Shannon and Khinchin, is based on four axioms, where the fourth one concerns additivity.
The group theoretic entropies make use of formal group theory to replace this axiom with a more general composability axiom.
As has been pointed out before, generalised entropies crucially depend on the number of allowed number degrees of freedom $N$.
The functional form of group entropies is restricted (though not uniquely determined) by assuming extensivity on the equal probability ensemble,
which leads to classes of functionals corresponding to sub-exponential, exponential or super-exponential dependence of the phase space volume $W$ on $N$.
We review the ensuing entropies, discuss the composability axiom, relate to the Gibbs' paradox discussion and explain why group entropies may be particularly relevant
from an information theoretic perspective.

\end{center}

\tableofcontents

\section{Introduction}

The aim of this paper is to discuss and review the construction of a class of entropies recently introduced, called group entropies.
We shall make several preliminary observations in order to ensure that our line of thinking is transparent.

In thermodynamics, according to Clausius the entropy $\Delta S = Q/T$ is defined macroscopically  by its change in terms of heat exchanged $Q$ and temperature $T$.
A connection to the microscopic world is obtained in statistical mechanics by Boltzmann's expression
\begin{equation}
	S[p] = -k_B\sum_i p_i\ln p_i=k_B\ln W.
	\label{Boltzmann-Shannon}
\end{equation}
where the last equality is valid on the equal probability ensemble $p_i=1/W$ where $p_i$ is the probabilistic weight of state $i$ and $W$ denotes the number of available states.

Jaynes made contact with information theory and pointed out that Boltzmann's micro canonical and canonical ensembles can be viewed as the probabilistic weights that maximise the Boltzmann-Shannon entropy functional in Eq. (\ref{Boltzmann-Shannon}) under suitable constraints. The micro canonical ensemble is obtained when only the normalisation constraint is imposed, whereas the canonical one arises when both normalisation and the average energy constraint are both assumed \cite{Jaynes1957}.

Here we will think of entropies in the spirit of information theory, i.e. functionals on probability space.
Therefore the first three S-K axioms\footnote{The four Shannon-Khinchin axioms \cite{Khinchin1957} assumes (1) S[p] to be a continuous function of the $p_i$, (2) that is maximized
by the uniform distribution $p_i=1/W$, (3) that the entropy is left unchanged if a  state of zero probability is added $S[p,0]=S[p]$ and (4) that the entropy is additive,
see e.g. \cite{Hanel_Thurner2011,Tempestaprepr2015}} are unobjectionable and the entropy of a system in the equal probability ensemble $p_i=1/W$ will be considered to a be a measure of uncertainty.
That said, it is natural to assume that entropy in the equal probability ensemble is extensive (in the limit of a large number of particles),  
i.e.  $S(p_i=1/W)=\lambda N$, $N\gg 1$. Namely, the more particles the more uncertain is the least biased ansatz $p_i=1/W$.
Hence we make extensivity in the sense clarified above a required property of the entropies we are going to consider. This is of course also done within the $q$-statistics framework
\cite{Tsallis2009}. It is also worth to recall that extensivity is a necessary condition for an entropy to play the role as a rate function in large deviation theory \cite{Ellis1985,Touchette2009}.

Once having established how the entropy of the entire system in the uniform ensemble scales with the number of particles (degrees of freedom), we need to make an equally important decision about composition of systems. Imagine a system that is obtained by merging two given systems $A$ and $B$ and imagine that $A$ and $B$ are statistically \textit{independent}. We start analysing this case not because we believe that real systems typically could be considered as collections of independent subsystems (although in classical thermodynamics it is often so); this is even less true when dealing with complex systems. Actually, the independent case serves two purposes. First, one can always formally consider independent systems as constituting a whole and the entropy needs to be defined so it can handle this. Secondly, this requirement allows for important mathematical constraints on the entropy and, as explained in Section 2,  establishes a link to group theory.

More precisely, since $A$ and $B$ are assumed to be independent, we can now either consider the cartesian product $A\times B$ of the states of the systems $A$ and $B$ as
one system and compute the entropy of $S(A\times B)$. Or we may as well first compute the entropy of the parts $S(A)$ and $S(B)$ and afterwards decide to consider $A\times B$ as a whole.
We recall that entropies are functionals on probability space, which define a probabilistic weight for each of the microstates of a given system. For the independent combination considered here,
 we of course have that microstates of $A\times B$ are given by the combined states $(i,j)$ where $i$ and $j$ refer to the specific microstates of $A$ and $B$, respectively.
The independence ensures that the probability distributions describing $A$, $B$ and $A\times B$ are related as $p_{i,j}^{A\times B}=p^A_ip^B_J$. So we need to have that the entropy functional
computed using $p_{i,j}^{A\times B}$ is consistent with the combination of the entropies obtained by computing the functional on $p_i^A$ and $p_j^B$ and then combine the result.
That is to say, we need a function $\phi(x,y)$ that takes care of the combination of the two independent systems $A$ and $B$ into one whole
\beq
S(A\times B)=\phi(S(A),S(B)).
\label{compo}
\eeq
This relation is of course basic in as much as it is a formality to consider the cartesian product $A\times B$ as a whole or as combined of two independent subsystems $A$ and $B$. The ultimate meaning of eq. \eqref{compo} is

Eq. (\ref{compo}) should therefore  be satisfied for all possible choices of $p_i^A$ and $p_j^B$. In Sec. 2 below we discuss the properties of $\phi(x,y)$ in more detail. Here we just mention that Eq. (\ref{compo}) ensures that in cases where the entire system can be considered to be a collection of subsystems, the entropy of a composed system $S(A \times B)$ depends on the entropy of A and the entropy of B  of the component systems only, without the need for a microscopic description of them. So, in this way one can associate naturally the notion of entropy to a macroscopic system starting from the knowledge of its constituents.

Complex systems are often defined as involving some degree of emergence, which is captured by the famous Aristoteles quote "The whole is greater than the sum of the parts".
A concrete explicit example of such situations was recently considered by introducing the so called pairing model \cite{PairingModel}
in which particles may combine to form new paired states which are entirely different from the single particle states.
For a specific example think of hydrogen atoms. When two hydrogen atoms combine to form a hydrogen molecule, $H_2$ bound states are formed which cannot be reach within the
cartesian product of the phase space of the two individual hydrogen atoms \cite{Hendry2010}. So when dealing with complex systems the independent combination of subsystems
will typically be different from the whole \cite{Morin2008}. Let us by $AB$ denote the system obtained by bringing the $N_A$ particles of system $A$ together with the $N_B$ particles of
system $B$ and allowing for the two sets of particles to establish all possible interactions or interdependencies between the particles from $A$ and the particles from $B$.
In the example of pairing $AB$  will include all the possible states of A combined as cartesian products with all the possible states of $B$ and in addition $AB$
will moreover contain all the states consisting of paired states between particles in $A$ and particles in $B$. Therefore  $AB\neq A\times B$   since the states of
$A\times B$ consists of the states that can be labelled as $(i,j)$ where $i=1,..,W_A$ runs through all the states of system $A$ and $j=1,...,W_B$ runs through all the
states of system $B$. New \textit{emergent} states formed by combining particles from $A$ and $B$ are not included in $S\times B$. To illustrate this think of system $A$ and
system $B$ as consisting of a single hydrogen atom each. Then $A\times B$ is the set $A\times B=\{({\bf r}_a,{\bf p}_a),({\bf r}_b,{\bf p}_b)\}$ where ${\bf r}_i$ and ${\bf p}_i$ with $i=a,b$ are
the position and momenta of the hydrogen atom $A$ or $B$. The combined system $AB$ in contrast contains all the states of $A\times B$ in addition to the new emergent molecular
states $H_2$ consisting of the $A$ hydrogen atom bound together with the $B$ hydrogen atom.

We require that the entropy evaluated on the equal probability ensemble for the fully interacting system satisfies (asymptotically) extensivity, i.e. that
\beq	
S(AB) = S(1/W(AB)) \simeq  \lambda N_{AB} = \lambda (N_A+N_B)\; {\rm for} \;N_A\gg 1\; {\rm and }\; N_B\gg 1.
\label{ent_of_AB}
\eeq
But we cannot in general insist that
\beq
S(AB)=\phi(S(A),S(B))=S(A\times B).
\label{AB_versus_AxB}
\eeq
Eq. (\ref{AB_versus_AxB}) can only be satisfied in the Boltzmann-Shannon case for which the entropy is additive and $\phi(x,y)=x+y$. Below we'll discuss in detail how the functional dependence
$W(N)$ of the total number of states on the number of particles will determine the properties of the entropy. Here we just note that when $AB=A\times B$ we need $W(AB)=W(A)W(B)$
and therefore $W(N)$ to be exponential in $N$, which is the Boltzmann-Shannon case. We'll see below that when $W(N)$ is different from the exponential,
one either gets entropies equivalent to the Tsallis entropies, for sub-exponential algebraic dependence $W(N)=N^a$, or new group entropies for super-exponential phase space
growth rates $W(N)=N^{\gamma N}$.

For complex systems, for which entropies don't add, the group entropies discussed here immediately suggest a measure of how complex a system is. Precisely because $A\times B\neq AB$ for complex systems
 and therefore the entropy of the fully interdependent system $AB$ is different from the cartesian combination $A\times B$ a measure of the essential emergent interdependence can be constructed as
\beq
\Delta(AB) = S(A\times B)-S(AB)= \phi(S(A),S(B)) - S(AB).
\label{Delta}
\eeq
This measure can be thought of as a possible generalisation of the usual mutual information and could perhaps e.g. be useful as an alternative to
Tononi's  Integrated Information \cite{Tononi_2001,Tononi_2008}  as a measure that can quantify very entangled complex systems such as perhaps consciousness. For further discussion of this
complexity measure we refer to \cite{Tempesta_Jensen2018}.

The remainder of the article is organised as follow. In Section 2, we present a brief and self-consistent introduction to group entropies. We explain in  Section 3 how the phase space growth volume $W(N)$ determines the group law that in turn defines the entropy and the rule for composing statistically independent systems. We discuss the two families of entropies, the trace form and the non-trace form and explain that $\phi(x,y)$ has the same functional form for the two classes for a given form of $W(N)$. We relate the group entropies to existing entropies and discuss in Section 4 the probabilities $p_i$ derived by  maximising the entropy under constraints.

\section{Basic results on group entropies}
In this section, we shall present a brief introduction to some basic aspects of the theory of group entropies. The mathematical apparatus will be kept to a minimum. For an extensive discussion, the reader is referred to the original papers \cite{Tempesta2011}, \cite{Tempesta2016}, \cite{Tempesta2015}, \cite{ET2017}.
We start out with the composition requirement in Eq. (\ref{compo}). We need to require that $i)$ $\phi(x,y)= \phi(y,x)$, since $A$ and $B$ can obviously be interchanged.
At the same time, we also require that the process of composition can be made in a associative way: $ii)$ $\phi(x,\phi(y,z))=\phi(\phi(x,y),z)$.
Finally, if system $B$ is in a state of zero entropy, we wish that the entropy of the composed state $A\times B$ coincides with the entropy of $A$. In other words, $iii)$ $\Phi(x,0)=x$.
We shall say that an entropy satisfies the \textit{composability axiom} if there exists a function $\phi(x,y)$ such that Eq. (\ref{compo}) is satisfied, jointly with the previous properties of
commutativity, associativity and composition with a zero-entropy state \cite{Tempestaprepr2015}, \cite{Tempesta2016}.

In order to ascertain the plausibility of the composability axiom, observe that, first of all, it is satisfied by Boltzmann's entropy. It is a crucial requirement for possible thermodynamical applications. Indeed, it means that the entropy of a (independently) composed system depends on the macroscopical configuration of the component systems only, without the need for a microscopic description of them. So, in this way one can associate naturally the notion of entropy to a macroscopic system starting from the knowledge of its constituents. At the same time, property \eqref{compo} is related to Einstein's likelihood principle \cite{ST2016}.

From a mathematical point of view, the composability axiom is equivalent to the requirement that $\phi(x,y)$ is a group law in the sense of formal group theory \cite{Haze}.
This is the origin of the group theoretical structure associated with the class of generalised entropies called group entropies \cite{Tempesta2011}, \cite{Tempesta2016}, \cite{Tempestaprepr2015}.
To be precise, a \textit{group entropy} is an entropic function satisfying the first three Shannon-Khinchin axioms and the composability axiom for all possible probability distributions.
This is called composible in a strong sense. If an entropy is only composable on the uniform distribution, one talks about weak composability.

The main connection between generalised entropies and group theory is the composability axiom. An interesting aspect is that the study of the algebraic structure defined by this set of requirements
has been developed in a completely different context, namely algebraic topology, in the second half of the past century. Here all we need is that a
\textit{one-dimensional formal group law} over a ring $R$ \cite{Haze} is a formal power series in two variables of the form
\begin{equation}\label{phi}
\phi(x,y)=x+y+\sum_{ij}a_{ij}x^{i}y^{j}
\end{equation}
that satisfies the properties $i)$--$iii)$. The theory of formal groups was introduced by Bochner in the seminal paper \cite{Bochner1946} and developed in algebraic topology, analysis,
and other branches of pure and applied mathematics by G. Faltings, S. P. Novikov,  D. Quillen, J. P. Serre and many others \cite{Haze}, \cite{Bukhshtaber1971}. For recent
applications in number theory, see also \cite{TANN2010}, \cite{TTAMS2015}.


A property crucial for the subsequent discussion is the following: given a one-dimensional formal group law $\phi(x,y)$, there exist a series $G(t)=t+\sum_{k=2}^\infty\beta_kt^{t}$ such that
\beq\label{G}
\phi(x,y)= G(G^{-1}(x)+G^{-1}(y)).
\eeq
The relation between group entropies and formal group laws is therefore immediate. A group entropy possesses a group law associated with it, expressed by a suitable function $\phi(x,y)$ of the
form \eqref{phi} which is responsible for the composition process for any possible choice of the probability distributions on $A$ and $B$. A natural question is how to classify group entropies.
To this aim, we recall that, generally speaking, we can distinguish between two large classes of entropy functions, the trace-form class and the non-trace-form one. In the first case, we shall deal
with entropies that can be written as $S= \sum_{i} f(p_i)$ for a suitable one-variable function $f(x)$. The prototype of this family is Boltzmann's entropy. If an entropy does not possess this form,
it is said to be of non-trace-form. The most well-known example of a non-trace form entropy is R\'enyi's entropy. In this paper we shall focus on the following realisations of the two classes.

For the {\it trace form class}, we shall consider the general functional \cite{Tempesta2016} \\
\beq
S[p] = \sum_{i=1}^{W} p_i G\bigg(\ln\frac{1}{p_i}\bigg).
\label{trace}
\eeq
called the universal-group entropy (since it is related with the algebraic structure called Lazard's universal formal group). Here $G(t)$ is an arbitrary real analytic
invertible function such that $G(0)=0$, $G'(0)=1$. \\
For the \noindent {\it non-trace form class} we shall consider the functional \cite{Tempesta2016}\\
\beq
S[p]=\frac{G\left(\ln(\sum_{i=1}^{W} p_i^\alpha)\right)}{1-\alpha}.
\label{non-trace}
\eeq
that has been called $Z$-entropy. Both families of entropies are assumed to satisfy the first three Shannon-Khinchin axioms for suitable choices of $G(t)$.
The main difference between the trace form and the non-trace form class is encoded in a theorem proved in \cite{ET2017}, where it was shown that the most general trace-form entropy satisfying
Eq. (\ref{compo}) is Tsallis entropy, with Botzmann's entropy as an important special case. The infinitely many other trace-form entropies only fulfil the composition law Eq. (\ref{compo})
on the restricted part of probability space consisting of uniform probabilities $p_{i,j}^{A\times B}=1/W_{A\times B}=1/(W_AW_B)=p_i^A p_j^B$. Therefore, these entropies are said to be weakly
composable \cite{Tempesta2016}. Instead, the non-trace form entropy \eqref{non-trace} is composable for any combination $A\times B$ of systems $A$ and $B$ with $p_{i,j}^{A\times B}=p^A_ip^B_j$.

\section{From phase space volume to group entropies}
Extensivity and the dependence on the size of phase space have often played a role in the analysis of entropies. For the case of Tsallis entropy, the requirement of
extensivity is used to determine the value of the  parameter $q$ \cite{Tsallis2009,Tsallis15377}; the importance of the dependence of the entropy on the available number of micro state $W$ was discussed
in \cite{Hanel_Thurner2011}. Here we describe how exploiting the relation between the number of micro states $W$ and the number of particles (or degrees of freedom) $N$ allows  one
to find the functional form of the group entropies, see \cite{PairingModel,Tempesta_Jensen2018}. For a discussion not assuming the composability requirement and hence the group structure
see \cite{Hanel2011,Korbel_Hanel_Thurner2018}.

We consider how the group-theoretic entropies deal with the three asymptotic dependencies of the phase space volume
\begin{itemize}
	\item[(I)] Algebraic \hspace{1.7cm} $W(N)=N^a$ \hspace{.8cm} with \hspace{.5cm} $W^{-1}(t)=t^{\frac{1}{a}}$,
	\item[(II)] Exponential \hspace{1.35cm}  $W(N)=k^N$ \hspace{.8cm} with  \hspace{.5cm} $W^{-1}(t)=\frac{\ln t}{\ln k}$,
	\item[(III)] Super-exponential \hspace{0.25cm}$W(N)=N^{\gamma N}$ \hspace{.5cm} with  \hspace{.5cm} $W^{-1}(t)=\exp[L(\frac{\ln t}{\gamma})]$.
\end{itemize}
Here $L(t)$ denotes the Lambert function.

Now we shall discuss how the extensivity requirement and the functional form of $W(N)$ determine the function $G(t)$,
which in turn determines the entropy according to formulae \eqref{trace}, \eqref{non-trace}.

Before we enter into technical details let us clarify how the present theory relate to previous investigation.

First, what could make the entropy non-additive? For the exponential case (I) we will find that the composition in Eq. \ref{compo} corresponds to simple addition $\phi(x,y)=x+y$.
This is the traditional Boltzmann-Shannon case. All four S-K axioms, including the 4th additivity axiom, are satisfied and in accordance with the uniqueness theorem \cite{Khinchin1957}
we find $S[p]=-\sum_i p_i\ln p_i$.  So, as one could expect, an exponential-type phase space volume is related with
additivity and no essential emergence of interdependence amongst the components of the considered system.  The situation turns out to be different for the cases (I) and (III) above.
In both these cases $W(AB)\neq W(A\times B)=W(A)W(B)$. In the sub-exponential case (I) the fully interdependent system $AB$ has fewer states available than $W(A)W(B)$. This situation is akin to how the Pauli principle prevents a set of Fermions  from occupying all possible combinations of single particle states.   (III) the system $AB$ has more states available than $W(A)W(B)$, new collective states have emerged when $A$ and $B$ are combined\cite{PairingModel}.

Lieb and Yngvason has argued \cite{Lieb_Yngvason_2001} that from standard classical thermodynamics, with out any use of statistical mechanics, it follows that entropy must be additive and extensive.
We recall that the fourth Shannon-Khichin \cite{Tempestaprepr2015} axiom assumes additivity and since the four SK axioms uniquely lead to the Boltzmann-Shannon functional form, hence
we can only be consistent with traditional thermodynamics if we remain within the Shannon-Khinchin axiomatic framework. This implies that only case (II) $W(N)=k^N$ is consistent with
traditional thermodynamics. The two cases (I) $W(N)=N^a$ and $W(N)=N^{\gamma N}$ turns out not to be consistent with additivity, which takes one outside the framework of Boltzmann-Shannon-Khinchin
and therefore in accordance with Lieb and Yngvason outside standard thermodynamics:
\footnote{Mathematical structures for generalised thermodynamics have been considered by several authors, see  e.g. Tsallis\cite{Tsallis2009} and Naudts\cite{Naudts2011}
and may of course be of relevance to various systems, but to the extend they consider non-additive entropies they seem not to relate to standard thermodynamics of, say, heat flow and steam engines,
but probably rather to information theory.} i.e. we are naturally lead to the abstract conceptual framework of information theory.
We wish to stress that group entropies represent measures of complexity by information geometric means \cite{RRT2018} and can characterize limiting probability distributions  by means of a
maximum entropy procedure for systems where interdependence among its components makes $W(N)$ deviate from the exponential form.

Stepping outside the SK framework can of course be done in multiple ways. One may simply decide to entirely give up on the 4th axiom and only assume the first three.
The approach was considered in \cite{Hanel_Thurner2011}; the authors then study how to determine the functional form of the entropy from scaling properties of the phase space volume $S\sim f(W)$,
without reference to how $W$ relates to $N$. The group theoretic approach described here is of course related, but it requires that a entropy must be defined in a way that allows the
computation of the entropy of the independent combination $A\times B$ to be related in a consistent and unique way to the entropy of the parts $A$ an $B$.

\subsection{From $W(N)$ to $G(t)$}
We start from the requirement that the group entropy is extensive on the equal probability ensemble $p_i=1/W$, i.e. we require asymptotically for large $N$, and therefore large $W$ that
\beq
S\left(p_i=\frac{1}{W}\right) = \lambda N.
\label{extensive}
\eeq
We now consider separately the trace form case  (\ref{trace}) and the non-trace form  (\ref{non-trace}) one. For the first case,
\beq
S\left(\frac{1}{W}\right)=\sum_{i=1}^W\frac{1}{W}G(\ln(W))=\lambda N,
\eeq
Inverting the relation between $S$ and $G$ we obtain
\beq
G(t)=\lambda W^{-1}[\exp(t)].
\label{G(t)_asym}
\eeq
This is a consequence of the asymptotic extensivity. But we also need $G(t)$ to generate a group law, which requires $G(0)=0$ \cite{Tempesta2016,Tempestaprepr2015},
so we adjust the expression for $G(t)$ in Eq. (\ref{G(t)_asym}) accordingly and conclude
\beq
G(t)=\lambda\{W^{-1}[\exp(t)]-W^{-1}(1)\}.
\label{grp_law_trace}
\eeq
Assuming the non-trace form in Eq. (\ref{non-trace}) when inverting Eq. (\ref{extensive}), and ensuring $G(0)=0$ leads to
\beq
G(t)=\lambda(1-\alpha)\{W^{-1}[\exp(\frac{t}{1-\alpha})]-W^{-1}(1)\}.
\label{grp_law_non-trace}
\eeq
From the expressions (\ref{grp_law_trace}) and (\ref{grp_law_non-trace}) we can now list the entropies corresponding to the three classes (I), (II) and (III) of phase space growth rates.
A straight forward calculation gives the following results:\\

\noindent {\it Trace-form case}\\
\begin{itemize}
	\item[(I)] Algebraic, $W(N)=N^a$:
	\begin{eqnarray}
	S[p]&=\lambda \sum_{i=1}^{W(N)} p_i\left[(\frac{1}{p_i})^\frac{1}{a}-1\right]\\
	    &= \frac{1}{q-1}(1-\sum_{i=1}^{W(N)} p_i^q).
	\label{I-trace}
	\end{eqnarray}
	To emphasize the relation with the Tsallis q-entropy, we have introduced $q=1-1/a$ and $\lambda = 1/(1-q)$. Note that the parameter $q$ is determined by the exponent $a$, so it is
	controlled entirely by $W(N)$.
		
	\item[(II)] Exponential, $W(N) = k^N$:
	\beq
	S[p]=\frac{\lambda}{\ln k}\sum_{i=1}^{W(N)} p_i\ln \frac{1}{p_i}.
	\label{II-trace}
	\eeq
	This is of course the Boltzmann-Gibbs case.
	
	\item[(III)] Super-exponential, $W(N)=N^{\gamma N}$:
	\beq
	S[p]=\lambda \sum_{i=1}^{W(N)} p_i\left\{\exp\left[ L(-\frac{\ln p_i}{\gamma})\right]-1\right\}.
	\label{III-trace}
	\eeq

\end{itemize}

\noindent {\it Non-trace form case}\\
\begin{itemize}
	\item[(I)] Algebraic, $W(N)=N^a$:
	\beq
	S[p]=\lambda \left\{\exp\left[\frac{\ln(\sum_{i=1}^{W(N)} p_i^\alpha)}{a(1-\alpha)}\right]-1\right\}.
	\label{I-non-trace}
	\eeq
	
	\item[(II)] Exponential, $W(N) = k^N$:
	\beq
	S[p]=\frac{\lambda}{\ln k}\frac{\ln( \sum_{i=1}^{W(N)} p_i^\alpha)}{1-\alpha}.
	\label{II-non-trace}
	\eeq
    This is of course the R{\'e}nyi entropy.
	
	\item[(III)] Super-exponential,  $W(N)=N^{\gamma N}$:
	\beq
	S[p]=\lambda\left\{\exp\left[L\Big(\frac{\ln \sum_{i=1}^{W(N)}p_i^\alpha}{\gamma(1-\alpha)}\Big)\right]-1\right\}.
	\label{III-non-trace}
	\eeq
	This entropy was recently studied in relation to a simple model in which the components can form emergent paired states in addition to the combination of single particle states \cite{PairingModel}.
\end{itemize}
\subsection{The composition law $\phi(x,y)$}
We now derive the composition law introduced in Eq. (\ref{compo}) above. The composition is given in terms of the function $G(t)$ as in \cite{Tempesta2015,Tempestaprepr2015} according to the relations \\
\noindent {\it Trace-form case}\\
\beq
\phi(x,y)=G[G^{-1}(x)+G^{-1}(y)].
\label{trace_compo_G}
\eeq
\noindent {\it Non-trace form case}\\
\beq
\phi(x,y)=\frac{1}{1-\alpha}G[G^{-1}((1-\alpha)x)+G^{-1}((1-\alpha)y)].
\label{non-trace_compo_G}
\eeq
When we express $\phi(x,y)$ directly in terms of the phase space volume $W(N)$ by use of Eqns. (\ref{grp_law_trace}) and (\ref{grp_law_non-trace}) we arrive at the following expression valid for both trace and non-trace forms
\beq
\phi(x,y)=\lambda\left\{
W^{-1}\Big[
W(\frac{x}{\lambda}+W^{-1}(1))W(\frac{y}{\lambda}+W^{-1}(1))
\Big]
-W^{-1}(1).
\right\}
\label{phi_W}
\eeq

The specific expressions for $\phi(x,y)$ for the three phase space growth rates are obtained from Eq. (\ref{phi_W}) by substituting the appropriate expressions for $W(N)$ and $W^{-1}(t)$ given by
\begin{itemize}
	\item[(I)] Algebraic, $W(N)=N^a$:
	\beq
	\phi(x,y) = x+y+\frac{1}{\lambda}xy=x+y+(1-q)xy.
	\eeq
	The case of Tsallis entropy.
	
	\item[(II)] Exponential, $W(N) = k^N$:
	\beq
	\phi(x,y)=x+y.
	\eeq
	The Boltzmann and R{\'e}nyi case.
	
	\item[(III)] Super-exponential, $W(N)=N^{\gamma N}$:
	\beq
	\phi(x,y)=\lambda\left\{
	\exp\Big[
	L\big(
	(1+\frac{x}{\lambda}\ln(1+\frac{x}{\lambda})+(1+\frac{y}{\lambda}\ln(1+\frac{y}{\lambda})
	\big)
	\Big]-1\right\}
	\eeq
	For examples of models relevant to this growth rate and composition law see \cite{PairingModel,Korbel_Hanel_Thurner2018}.
\end{itemize}

\section{Maximum entropy ensembles}
Let us now consider the probabilities derived from the group entropies by maximizing them under very simple constraints.
As usual, we shall introduce the constraints by means of Lagrange multiplies and analyze the functional
\beq
J[p]=S[p] -\sum_{n=1}^M\lambda_n g_n[p],
\eeq
for $M$ constraints given by the functionals $g_n[p]$. Traditionally, one uses the first constraint to control the normalization
\beq
g_1[p] = \sum_{i=1}^W p_i -1
\eeq
and the second one to determine the average of some observable $E$. In physics, this observable is typically the average of the systems energy $\bar{E}=\langle E\rangle -E_0$ measured
from the ground state level $E_0$
\beq
g_2[p] = \sum_{i=1}^W (p_i(E_i-E_0) - \bar{E}).
\eeq
With these two constraints, from the extremal condition $\delta J/\delta p_i =0$ we obtain
\beq
\frac{\delta S[p]}{\delta p_i}= \lambda_1 + \lambda_2(\Delta E_i-\bar{E}).
\label{Extrem_Eq}
\eeq
Here $\Delta E_i=E_i-E_0$.\\
\subsection{Trace form entropies}
The derivatives of $S[p]$ for the three trace forms Eqns. (\ref{I-trace}), (\ref{II-trace}) and (\ref{III-trace}) are\\

\begin{itemize}
\item[(I)] Algebraic -- $W(N)=N^a$:
\beq\frac{\delta S}{\delta p_i} = \lambda (1-\frac{1}{a})\left(\frac{1}{p_i}\right)^\frac{1}{a}.
\label{I-trace-deriv}
\eeq

\item[(II)] Exponential -- $W(N)=k^N$:
\beq
\frac{\delta S}{\delta p_i} =\frac{\lambda}{\ln k}(\ln\frac{1}{p_i}-1).
\label{II-trace-deriv}
\eeq

\item[(III)] Super-exponential -- $W(N)=N^{\gamma N}$:

\beq
\frac{\delta S}{\delta p_i} =\lambda\left\{ \exp\big[ L(X_i)\big]-\frac{1}{\gamma}\frac{1}{1+L(X_i)}-1\right\}, \;\;{\rm with}\;\; X_i=\frac{1}{\gamma}\ln\frac{1}{p_i}.
\label{III-trace-deriv}
\eeq
\end{itemize}
The functional form of $\delta S/\delta p_i$ in Eq. (\ref{I-trace-deriv}) and Eq. (\ref{II-trace-deriv}) allow solving Eq. (\ref{Extrem_Eq}) to express $p_i$ as
\begin{itemize}
\item[(I)] Algebraic -- $W(N)=N^a$:
\beq
p_i = \frac{[1+\beta (\Delta E_i-\bar{E})]^{-a}}{Z},
\label{I-trace_p_i}
\eeq
where formally $\beta =\lambda_2/\lambda_1$ and $Z=\sum_i [1+\beta (\Delta E_i-\bar{E})]^{-a}$. These probabilistic weights correspond to Tsallis' q-exponentials.	
\item[(II)] Exponential -- $W(N)=k^N$:
\beq
p_i=\frac{\exp[-\beta(\Delta E_i-\bar{E}]}{Z}
\label{II-trace_p_i}
\eeq
where formally $\beta =\lambda_2/\lambda_1$ and $Z=\sum_i 	\exp[-\beta(\Delta E_i-\bar{E})]$. As expected, we have re-derived the Boltzmann weights
starting from the trace form of the group entropies and exponential phase space growth rates.
\end{itemize}
The transcendental nature of the expression for $\delta S/\delta p_i$ in Eq. (\ref{III-trace-deriv}) seems to prevent one from deriving
a closed-form expression for $P_i$ in the case of super-exponential phase space growth rate $W(N)=N^{\gamma N}$, having assumed the trace-form (weakly composable) expression \eqref{III-trace}
for the entropy. We shall see below that the situation is different when starting from the non-trace forms.

\subsection{Non-trace form entropies}
The form of the entropies for the non-trace case given in Eqns. (\ref{I-non-trace}),
(\ref{II-non-trace}) and (\ref{III-non-trace}) all lead to the same functional expression as when starting from the trace form in the algebraic case expression, Eq. (\ref{I-trace_p_i}, namely
\beq
p_i = \frac{[1+\beta (\Delta E_i-\bar{E})]^{\frac{1}{1-\alpha}}}{Z}
\eeq
where formally $\beta =\lambda_2/\lambda_1$ and $Z=\sum_i [1+\beta (\Delta E_i-\bar{E})]^{\frac{1}{1-\alpha}}$.  This expression for $p_i$ is reminiscent of the Tsallis q-exponential.


\section{Discussion}
We have seen that the group theoretic entropies offers a systematic classification of entropies according to how the phase space growths with number of particles,
or number of degrees of freedom. The formalism allows for a systematic generalisation of a statistical mechanics description to non-exponential phase spaces and reduces to the Boltzmann-Gibbs
case when the $W(N)$ is exponential. A new measure of complexity as function of system size follows right away.

We wish to point out that group entropies are an interesting tool in information geometry, since they can be used to define Riemannian structures in statistical spaces via suitable divergences
(or relative entropies)  \cite{Amari2016} associated with them. This has been proved in  \cite{RRT2018} for Z-entropies and in  \cite{GBP2018} for the universal-group entropy.
Also, a quantum version of these entropies can be used as an entanglement measure for spin chains \cite{Tempestaprepr2015}. Work is in progress along these lines.

\end{document}